\newlength\smallfigwidth
\documentclass[prb,aps,twocolumn,superscriptaddress,floatfix,showpacs]{revtex4-1}
\usepackage{bm,textcomp}

\usepackage{color}
\usepackage{amsmath}
\usepackage{graphicx}
\usepackage{graphics}
\usepackage{pstricks}
\usepackage{pdftexcmds}
\usepackage{ifpdf}
\usepackage{amssymb}
\usepackage{tikz}
\usepackage{multirow}
\makeatletter
\@ifundefined{textcolor}{}

{%
 \definecolor{BLACK}{gray}{0}
 \definecolor{WHITE}{gray}{1}
 \definecolor{RED}{rgb}{1,0,0}
 \definecolor{GREEN}{rgb}{0,1,0}
 \definecolor{BLUE}{rgb}{0,0,1}
 \definecolor{CYAN}{cmyk}{1,0,0,0}
 \definecolor{MAGENTA}{cmyk}{0,1,0,0}
 \definecolor{YELLOW}{cmyk}{0,0,1,0}
 }
 
\definecolor{J1}{rgb}{0.0, 0.5, 1.0}
\definecolor{Jx}{rgb}{0.6, 0.33, 0.73}
\definecolor{Jp}{rgb}{0.28, 0.24, 0.2}
\definecolor{bond}{rgb}{0.2, 0.2, 0.9}
\allowdisplaybreaks
\makeatother
\def\vr {{\bf r}}

\def \ba {\begin{eqnarray}}
\def \ea {\end{eqnarray}}

\begin{document}

\title{Phase transitions, order by disorder and finite entropy in the  Ising  antiferromagnetic bilayer honeycomb lattice}

\author{F.A. G\'omez Albarrac\'in}
\email[]{albarrac@fisica.unlp.edu.ar}
\altaffiliation{Current address: Instituto de F\'isica de L\'iquidos y Sistemas Biol\'ogicos (IFLYSIB), UNLP-CONICET, La Plata 1900, Argentina and Departamento de F\'isica, Facultad de Ciencias Exactas, Universidad Nacional de La Plata, c.c. 16, suc. 4, La Plata 1900, Argentina}
\author{H.D. Rosales}
\affiliation{Instituto de F\'isica de La Plata -CONICET. Facultad de Ciencias Exactas, Universidad Nacional de La Plata, C.C. 67, 1900 La Plata, Argentina and Departamento de F\'{i}sica, FCE, UNLP, La Plata, Argentina}
\altaffiliation{Current address: Instituto de F\'isica de L\'iquidos y Sistemas Biol\'ogicos (IFLYSIB), UNLP-CONICET, La Plata 1900, Argentina and Departamento de F\'isica, Facultad de Ciencias Exactas, Universidad Nacional de La Plata, c.c. 16, suc. 4, La Plata 1900, Argentina; Facultad de Ingenier\'ia, UNLP, La Plata, Argentina}
\author{Pablo Serra}
\affiliation{Facultad de Matem\'atica, Astronom\'ia, F\'isica y Computaci\'on, Universidad Nacional de C\'ordoba and IFEG-CONICET,
Ciudad Universitaria, X5016LAE C\'ordoba, Argentina}
\begin{abstract}
We present an analytical and numerical study of the Ising model on a bilayer honeycomb lattice including interlayer frustration
and coupling with an external magnetic field.  
First, we discuss  the exact $T=0$ phase diagram, where we find finite entropy phases for different magnetisations.
Then, we study the magnetic properties of the system at finite temperature using complementary analytical techniques  (Bethe lattice), 
and two types of Monte-Carlo algorithms (Metropolis and Wang-Landau).
We characterize the phase transitions and discuss the phase diagrams. 
The system presents a rich phenomenology: there are first and second order  transitions, low-temperature phases with extensive degeneracy,
and order-by-disorder state selection.
\end{abstract}
\maketitle{}

\section{Introduction}
\label{sec:intro}
 The continuous exploration of frustrated spin systems in the last years has been
 driven by the role of frustration 
to induce unconventional magnetic orders or macroscopic degeneracy in the ground state with no long-range ordering [\onlinecite{Diep}].
However,  this macroscopic degeneracy will depend critically of the coordination number  and/or the spin representation. 
For instance, on one hand in the antiferromagnetic (AF) triangular lattice the classical Heisenberg model has a unique ordered ground state [\onlinecite{ZitoTrig}] 
while the classical Ising model shows a large degeneracy in the ground state [\onlinecite{Wannier}]. On the other hand, in the AF kagome lattice,
both models (Heisenberg and Ising)   present a disordered ground state with macroscopic degeneracy [\onlinecite{OBDkagome}].
The interaction with a magnetic field lowers the symmetries of these systems, and may lead to a total or partial reduction of the ground state 
degeneracy.

In the case of the  honeycomb lattice, since it is bipartite, the AF model with nearest-neighbor interactions is not frustrated. Additional interaction terms, for example next-nearest neighbors, 
are needed to introduce magnetic frustration [\onlinecite{GaneshJ1J2}-\onlinecite{RosalesJ1J2}]. 
In the last few years, several works [\onlinecite{Ganesh2011, Kandpal2011,Matsuda2010,Lamas2016,Richter2017,AbInitio2017,Bishop2017}] 
have been published motivated by non-trivial phases found in the material Bi$_3$Mn$_4$O$_{12}$(NO$_3$) [\onlinecite{Smirnova2009}].
Experimental evidence shows that this material can be modeled as a weakly coupled bilayer honeycomb lattice where magnetic frustration, suggested by the large negative value of the Curie-Weiss
temperature $\Theta_{CW}= -257$K, could play an important role in low-temperature properties. For that reason, in a recent paper we studied the antiferromagnetic 
bilayer honeycomb lattice in the highly frustrated case for classical spins [\onlinecite{Albarracin16}]. Frustration in this model 
is given by a competition between intralayer nearest-neighbors, and two interlayer couplings (all antiferromagnetic). In that work,  we found that 
 due to the the high level of frustration,  an external  magnetic field induced  the selection of non trivial low temperature phases.

In this work, we study the Ising model in  the honeycomb bilayer lattice and 
explore the  magnetic properties for the full range of the antiferromagnetic couplings. In order to do this, we resort
to a combination of analytical and numerical techniques (Bethe lattice approximation, Metropolis and Wang-Landau Monte-Carlo simulations).
Surprisingly, we find a very good agreement between the numerical results and the mean field technique. As we will show in the following sections, the interplay between these methods
has proved essential for a thorough study of the model. 

The paper is structured as follows: first, we introduce the model and present the $T=0$ phase diagrams
in Sec. \ref{sec:PhaseDiagramT0}. In Sec. \ref{sec:methods} we present the methods and  the order parameters used to study 
the low temperature behavior of the system, discussing the characteristics, advantages and limitations of each technique.
In Sec. \ref{sec:results} we present the low temperature phase diagrams for different regimes of the model.
We find different types of phase transitions,  highly degenerate phases and selection of states by thermal fluctuations.
Concluding remarks are presented in Sec. \ref{sec:Conclusions}.

\section{Model and $T=0$ Phase Diagram}
\label{sec:PhaseDiagramT0}
\begin{figure} [h!] 
\includegraphics[width=0.8\columnwidth]{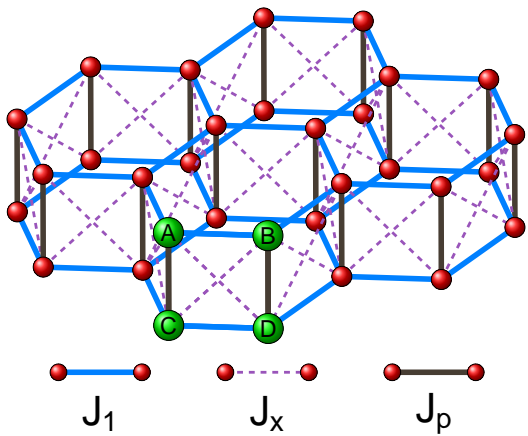}
\caption{\label{fig:model} (Color online) Bilayer honeycomb lattice. The (antiferromagnetic) couplings are indicated by different colors.
 $A,B,C,D$ label spins in a unit cell.  The interlayer coupling $J_p$ is indicated by (brown) vertical lines (joining sites $A-C$ and $B-D$),
 the intralayer coupling $J_1$  by (blue) horizontal ones (joining sites $A-B$ and $C-D$), and intralayer frustrating coupling $J_x$ is drawn as (violet) dashed lines
 (joining sites $A-D$ and $B-C$)}. 
\end{figure}

Let us define  the Ising model on  the antiferromagnetic bilayer honeycomb lattice as:

\begin{eqnarray}
\mathcal{H}=J_p\!\sum_{\vr}\left(\sigma^{A}_{\vr}\sigma^{C}_{\vr}+\sigma^{B}_{\vr}\sigma^{D}_{\vr}\right)-h\sum_{\vr,i}\sigma^{i}_{\vr}  \nonumber \\
 +\!\sum_{\langle\vr,\vr'\rangle}J_{1}\left(\sigma^{A}_{\vr}\sigma^{B}_{\vr'} + \sigma^{C}_{\vr}\sigma^{D}_{\vr'}\right) +J_{x}\left(\sigma^{A}_{\vr}\sigma^{D}_{\vr'} + \sigma^{B}_{\vr}\sigma^{C}_{\vr'} \right)
\label{eq:Hamiltonian}
\end{eqnarray}
\noindent where $\vr$  runs over unit cells,  $\langle\vr,\vr'\rangle$  denotes interactions within the cell and between  nearest-neighbor cells
and $i$ is the spin cell index $i={A,B,C,D}$ (with sites $A,B,C,D$ shown in Fig. \ref{fig:model}). The lattice structure and exchange antiferromagnetic 
couplings $J_1,J_x$ and $J_p$ are shown in Fig. \ref{fig:model}. $J_p$ is the coupling  joining the honeycomb layers (above and below), 
$J_{1},J_x$ are the nearest-neighbors in-plane and intraplane couplings respectively.
Note that the model in Eq. (\ref{eq:Hamiltonian}) can be mapped onto an identical one replacing
$J_1\leftrightarrow J_x$ by  exchange of opposite sites in each plaquette: $A\leftrightarrow C$ or $B\leftrightarrow D$. 
This $\mathcal{Z}_2$ symmetry will play an important role in the characterization of the low-temperature 
phases.

In the particular case of $h=0$ and $J_x=0$, the ground state has long-range N\'eel order composed of two opposite N\'eel states 
in each layer (for instance $\sigma_A=\sigma_D=-\sigma_B=-\sigma_C=+1$). 
A non zero value of $J_x>0$ introduces frustration, which leads to interesting phenomena even at zero temperature.
The magnetic properties of this model are controlled then by two factors  :the level of frustration 
and the magnetic field. 

We first study the zero temperature $T=0$ phase diagram. In order to do this, we need to describe the ground-state configurations 
for an individual plaquette. There are $16$ possible states in each 4-spin plaquette.
We consider only those with zero or positive magnetisation, and we are left with five types of plaquette arrangements. 
These  are listed in  Table \ref{table:plaq} [\onlinecite{foot1}] 
with their energy ($\mathcal{E}_0$), magnetisation 
($m_0=\frac{1}{4}(\sigma_A+\sigma_B+\sigma_C+\sigma_D)$) and degree of degeneracy ( $\mathcal{D}$).
We also introduce the notation for each configuration that we will use 
throughout this work: AF stands for antiferromagnetic ordering in each layer,
U for uniform order in a layer.

\begin{table} 
\begin{tabular}{|c|c| c| c|c|} 
\hline
&Notation & $\mathcal{E}_0$ & $m_0$  & $\mathcal{D}$\\
\hline
\begin{tikzpicture}
\draw [line width=1mm,color= J1] (0,0) -- (0.5,0);
\draw [line width=1mm,color= J1] (0,0.5) -- (0.5,0.5);
\draw [line width=1mm,color= Jp] (0,0) -- (0,0.5);
\draw [line width=1mm,color= Jp] (0.5,0) -- (0.5,0.5);
\draw [line width=0.7mm,color= Jx] (-0.1,-0.1) -- (0.6,0.6);
\draw [line width=0.7mm,color= Jx] (-0.1,0.6) -- (0.6,-0.1);
\shadedraw [ball color= green] (0,0) circle (0.15cm);
\shadedraw [ball color= green] (0.5,0) circle (0.15cm);
\shadedraw [ball color= green] (0.5,0.5) circle (0.15cm);
\shadedraw [ball color= green] (0,0.5) circle (0.15cm);
\node at (0,0) {{\large +}};
\node at (0.5,0) {{\large -}};
\node at (0.5,0.5) {{\large -}};
\node at (0,0.5) {{\large +}};
\end{tikzpicture}
&AF$_1$ & $2(J_p/3-J_1-J_x)$ & 0 & 2 \\
\begin{tikzpicture}
\draw [line width=1mm,color= J1] (0,0) -- (0.5,0);
\draw [line width=1mm,color= J1] (0,0.5) -- (0.5,0.5);
\draw [line width=1mm,color= Jp] (0,0) -- (0,0.5);
\draw [line width=1mm,color= Jp] (0.5,0) -- (0.5,0.5);
\draw [line width=0.7mm,color= Jx] (-0.1,-0.1) -- (0.6,0.6);
\draw [line width=0.7mm,color= Jx] (-0.1,0.6) -- (0.6,-0.1);
\shadedraw [ball color= green] (0,0) circle (0.15cm);
\shadedraw [ball color= green] (0.5,0) circle (0.15cm);
\shadedraw [ball color= green] (0.5,0.5) circle (0.15cm);
\shadedraw [ball color= green] (0,0.5) circle (0.15cm);
\node at (0,0) {{\large -}};
\node at (0.5,0) {{\large +}};
\node at (0.5,0.5) {{\large -}};
\node at (0,0.5) {{\large +}};
\end{tikzpicture}
&AF$_2$ & $2(-J_p/3-J_1+J_x)$ & 0 & 2\\
\begin{tikzpicture}
\draw [line width=1mm,color= J1] (0,0) -- (0.5,0);
\draw [line width=1mm,color= J1] (0,0.5) -- (0.5,0.5);
\draw [line width=1mm,color= Jp] (0,0) -- (0,0.5);
\draw [line width=1mm,color= Jp] (0.5,0) -- (0.5,0.5);
\draw [line width=0.7mm,color= Jx] (-0.1,-0.1) -- (0.6,0.6);
\draw [line width=0.7mm,color= Jx] (-0.1,0.6) -- (0.6,-0.1);
\shadedraw [ball color= green] (0,0) circle (0.15cm);
\shadedraw [ball color= green] (0.5,0) circle (0.15cm);
\shadedraw [ball color= green] (0.5,0.5) circle (0.15cm);
\shadedraw [ball color= green] (0,0.5) circle (0.15cm);
\node at (0,0) {{\large -}};
\node at (0.5,0) {{\large -}};
\node at (0.5,0.5) {{\large +}};
\node at (0,0.5) {{\large +}};
\end{tikzpicture}
&U$_2$ & $2(-J_p/3+J_1-J_x)$ & 0 & 2\\
\begin{tikzpicture}
\draw [line width=1mm,color= J1] (0,0) -- (0.5,0);
\draw [line width=1mm,color= J1] (0,0.5) -- (0.5,0.5);
\draw [line width=1mm,color= Jp] (0,0) -- (0,0.5);
\draw [line width=1mm,color= Jp] (0.5,0) -- (0.5,0.5);
\draw [line width=0.7mm,color= Jx] (-0.1,-0.1) -- (0.6,0.6);
\draw [line width=0.7mm,color= Jx] (-0.1,0.6) -- (0.6,-0.1);
\shadedraw [ball color= green] (0,0) circle (0.15cm);
\shadedraw [ball color= green] (0.5,0) circle (0.15cm);
\shadedraw [ball color= green] (0.5,0.5) circle (0.15cm);
\shadedraw [ball color= green] (0,0.5) circle (0.15cm);
\node at (0,0) {{\large -}};
\node at (0.5,0) {{\large +}};
\node at (0.5,0.5) {{\large +}};
\node at (0,0.5) {{\large +}};
\end{tikzpicture}
&UAF & $-2h/3$ & 1/2 & 4\\
\begin{tikzpicture}
\draw [line width=1mm,color= J1] (0,0) -- (0.5,0);
\draw [line width=1mm,color= J1] (0,0.5) -- (0.5,0.5);
\draw [line width=1mm,color= Jp] (0,0) -- (0,0.5);
\draw [line width=1mm,color= Jp] (0.5,0) -- (0.5,0.5);
\draw [line width=0.7mm,color= Jx] (-0.1,-0.1) -- (0.6,0.6);
\draw [line width=0.7mm,color= Jx] (-0.1,0.6) -- (0.6,-0.1);
\shadedraw [ball color= green] (0,0) circle (0.15cm);
\shadedraw [ball color= green] (0.5,0) circle (0.15cm);
\shadedraw [ball color= green] (0.5,0.5) circle (0.15cm);
\shadedraw [ball color= green] (0,0.5) circle (0.15cm);
\node at (0,0) {{\large +}};
\node at (0.5,0) {{\large +}};
\node at (0.5,0.5) {{\large +}};
\node at (0,0.5) {{\large +}};
\end{tikzpicture}
&U & $2(J_p/3+J_1+J_x-2h/3)$ & 1 & 1 \\
\hline
\end{tabular}
 \caption{\label{table:plaq} Plaquette configurations,
 energies ($\mathcal{E}_0$), magnetisation ($m_0$) and degree of degeneracy $\mathcal{D}$. }
\end{table}
Having constructed the different plaquettes configurations, listed in Table \ref{table:plaq}, we now discuss  the $T=0$ phase diagram for different cases:
\subsubsection{$h=0$ case}
\begin{figure} [h!]
 \includegraphics[width=1.0\columnwidth]{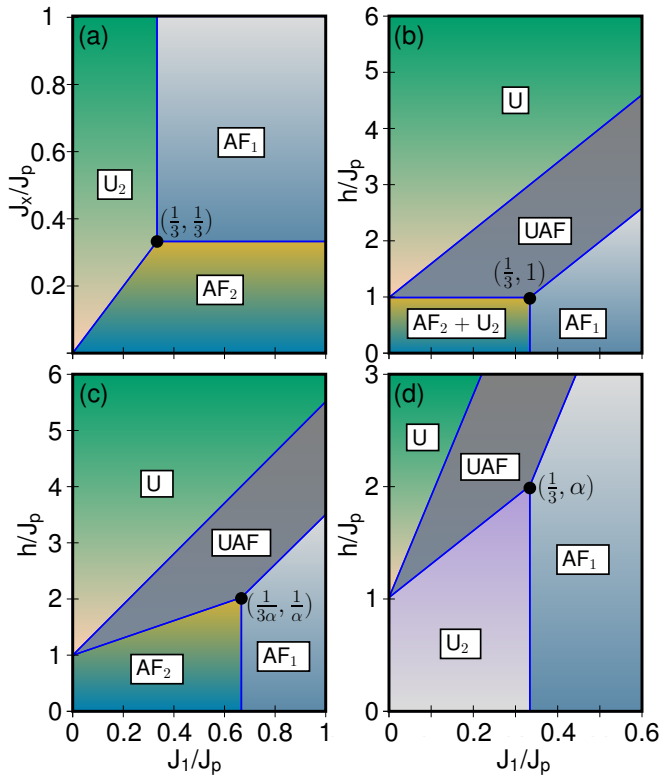}
 \caption{ \label{fig:phdiagT0}(Color online) $T=0$ phase diagrams.(a) $J_x$ vs $J_1$ in units of $J_p$ for $h=0$. $h/J_p$ vs 
 $J_1/J_p$ phase diagrams 
 for (b) $J_1=J_x$ (c) $J_x<J_1$ ($\alpha=J_x/J_1=1/2$) and (d) $J_x > J_1$ ($\alpha=J_x/J_1=2$)}
\end{figure}

We show the $J_x/J_p$ vs $J_1/J_p$ $T=0$ phase diagram of the model for the case of zero magnetic field in the upper-left panel of Fig. \ref{fig:phdiagT0}.  
Interesting features appear on the specific line $J_x=J_1<J_p/3$ where the ground state is highly degenerate because each plaquette can either be in a U$_2$ or AF$_2$ configuration.
The energy does not depend on $J_p$, and thus each $J_p$ pair can be flipped without energy cost. As a consequence the system has macroscopic degeneracy, 
and therefore a non zero entropy, at $T=0$.
The determination of the value of this entropy can be easily computed: it is the same as that of a random spin configuration in one of the layers
(the spins in the other layer are simply opposite). Therefore, the entropy per spin $s=S/N$ in units of $\ln 2$ (the value in the paramagnetic case) is $s=1/2$. 
In the highly frustrated point $J_1=J_x=J_p/3$, in addition to the degenerate configurations we just described there are two more possible states: 
all plaquettes with an AF$_1$ configuration. 

\subsubsection{$h>0$ case}
The effect  of an external field is summarized in  Fig. \ref{fig:phdiagT0} where we show the $h/J_p$ vs $T/J_p$ 
phase diagrams for three different regions of  interest for the  frustrating relation $J_x/J_1$: (b) $J_x/J_1=1$ (top right), (c) $J_x/J_1 < 1$ (bottom left) 
and  (d) $J_x/J_1 > 1$ (bottom right) characterized by the  total magnetisation $m$ defined as
\begin{equation}
m= \frac{1}{N}\sum_{I,\vr}\sigma^{I}_{\vr},\qquad I=A,B,C,D.
\label{eq:magnetisation}
\end{equation}
We found that there are three possible magnetisation plateaux: at $m=0$ (with structure U$_2$, AF$_1$ or AF$_2$), at $m=1/2$ (UAF)
and the saturation plateau $m=1$ (U) for $h>h_{sat}$. Explicit expressions for the critical fields are given in Appendix  \ref{AppendixA} .
As we just stated, the $m=1/2$ is given by UAF plaquettes. This configuration is highly degenerate:
one $J_p$ pair of the spins of a plaquette  has positive ($+$) spin, and the other one has two opposite spins. This pair has no
particular orientation, so in each unit cell there is one degree of freedom. Therefore, at this plateau there is finite entropy $s=1/4$.
For $J_1\neq J_x$, the $m=0$ plateau is simply doubly degenerate, and therefore has $s=0$ in the thermodynamic limit. The magnetic field then
takes the system from a $s=0$ phase to   a magnetised highly degenerate one, where the entropy is finite at $T=0$.

\section{Analytical and numerical approaches}
\label{sec:methods}

Having studied the zero temperature  phase diagram of the model presented in Sec.\ref{sec:PhaseDiagramT0}, 
in the next sections we look at the  effect of thermal fluctuations by means of three complementary methods, 
namely Bethe lattice approach and  Monte-Carlo simulations
(Metropolis and Wang-Landau), which we briefly describe in the following  subsections.

\subsection{Bethe lattice}
\label{sec:BetheLattice}

The Bethe lattice (BL) is a mean-field  approach that, for first neighbor interactions,
is equivalent to the Bethe approximation [\onlinecite{baxter}]. 
From  the point of view of correlations,  in a simple mean-field calculation no 
correlations are taken into account, while in Bethe lattice solutions 
short-range correlations are considered. In particular, for the model studied here,
the correlations between spins in a plaquette are taken into account in exact form.

\begin{figure}
\begin{center}
\includegraphics[width=0.3\textwidth]{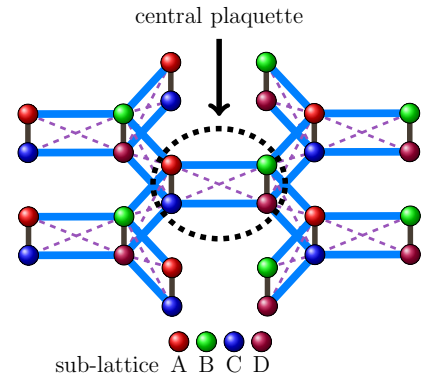}
\end{center}
\caption{\label{fig:fbl} (Color online) The $q=3$ bilayer Bethe lattice that reproduces the ground state
configurations of the bilayer honeycomb lattice.} 
\end{figure}

The Bethe lattice consists in the exact solution of a statistical model in the core of a Cayley tree.
In order to approximate  the bilayer honeycomb lattice, we define a bilayer Cayley 
tree, similar as the one used previously for different Ising-like  models by 
Hu {\it et al.} [\onlinecite{hio99}] and Albayrak and co-workers [\onlinecite{albayrak}], 
and for self-avoiding walks by Serra and Stilck [\onlinecite{serra14}].
The simplest  tree-like approximation for our model is a bilayer Cayley tree
with the same  coordination $q=3$ of the honeycomb lattice, and interlayer first and 
second interactions. It is important to note that the  model in this lattice
 reproduces the correct $T=0$ phase diagram (ground state) of the two-dimensional 
model described in Sec. \ref{sec:PhaseDiagramT0}.
In order to take statistical averages in the central zone of the Cayley tree 
(the Bethe lattice approach), we define it as
a central plaquette, as shown in Fig. \ref{fig:fbl}. 

The honeycomb lattice is bipartite, then, in the notation of Fig. \ref{fig:fbl},
the points $A$ and $C$ belong  to a sublattice $a$, and the points $B$ and $D$ to the
other sublattice $b$. In order to describe the different phases we need to define
eight partial partition functions (PPF) $Z^{a}_{\sigma_A\sigma_C}$,
and $Z^{b}_{\sigma_B\sigma_D}$
each one for the four possible values of the spins corresponding to a sublattice 
bilayer, $(1,1),(1,-1), (-1,1)$, and $(-1,-1)$. The tree-like structure of the lattice
allows us to write down   recursion relations (RR) for the  PPF.  
Once the RR are obtained,
the thermodynamic phases are   given by their stable fixed points. Where the 
stability line of two different fixed points are coincident, it represents a 
second order line.  For  first order lines we have to calculate the Bethe lattice 
free energy [\onlinecite{gujrati95,oss09}], and look for the line in the coexistence zone 
where the free energies of  both phases are equal. The PPF, free energy and other 
details of the calculations are developed in  Appendix B.

In general, the stability lines of the several  thermodynamic fixed points are given
as solutions of  sets of nonlinear coupled equations. However, for the important 
case  $h=0,\,J_1=J_x$, the paramagnetic stability line takes the simple form

\begin{equation} \label{eslh0j1jx}
e^{\frac{J_p}{T}}\,=\,\frac{\left(1+e^{4 \frac{J_x}{T}}\right) \sqrt{2 \left(
e^{8\frac{J_x}{T}}-3 \right)}}{e^{2 \frac{ J_x}{T}}  
\left(e^{4\frac{J_x}{T}}+3 \right)} \,
\end{equation}

\noindent The other curves in the phase diagram were calculated solving numerically
the set of exact algebraic coupled equations (see  Appendix B).

The order parameters and phase diagrams obtained within  the Bethe lattice approximation
are displayed in the following  subsections, which describe the MC techniques.

\subsection{Monte-Carlo simulations}
\label{sec:MC-simulations}

We simulated lattices with periodic boundary conditions in the two directions with Metropolis update[\onlinecite{Metropolis}] (MC-M) 
and Wang-Landau[\onlinecite{WangLandau01}] (MC-WL) methods. 
In both cases we used a single spin flip algorithm. We exploited the strength of each technique to fully understand the system, 
as commented below.

\subsubsection{Metropolis}
\label{sec:MC-Metropolis}

We performed MC-M simulations on lattices of $N=4\times L^2$ sites ($L=24-60$).  
To avoid the problem of low temperature ``freezing'' of the simulations, we used the
the annealing technique,
lowering the temperature as $T_{n+1}=0.9\times T_n$, from $T_i/J_p=5$ to $T_f/J_p \approx 0.1$.
We also averaged results over $100$ copies of the simulations,
generated from different random seeds. Data was taken in each copy averaging $4\times 10^5$ Monte-Carlo Steps (mcs),
after discarding $2\times 10^5$ mcs for thermalization. We measured the energy per spin, the magnetisation per spin (Eq. (\ref{eq:magnetisation})), the specific heat per spin 
\begin{equation}
C=\frac{\langle E \rangle^2-\langle E^2 \rangle}{N\,T^2},
\end{equation}
and three different order parameters to detect the three possible zero magnetisation arrangements of the plaquettes (see Table \ref{table:plaq}) defined as follows,
\begin{eqnarray}
\label{eq:OP1}
\text{OP}_{\text{AF}_1}=\frac{1}{N}\sum_{\vr}(\sigma^{A}_{\vr}+\sigma^{C}_{\vr}-\sigma^{B}_{\vr}-\sigma^{D}_{\vr}) \\
\label{eq:OP2}
\text{OP}_{\text{AF}_2}=\frac{1}{N}\sum_{\vr}(\sigma^{A}_{\vr}+\sigma^{D}_{\vr}-\sigma^{B}_{\vr}-\sigma^{C}_{\vr}) \\
\label{eq:OP3}
\text{OP}_{\text{U}_2}=\frac{1}{N}\sum_{\vr}(\sigma^{A}_{\vr}+\sigma^{B}_{\vr}-\sigma^{C}_{\vr}-\sigma^{D}_{\vr}) 
\end{eqnarray}
where $\vr$ runs over unit cells. With the previous definitions, the local order parameters take values $-1$ or $1$ when the plaquette is in the specific configuration 
(AF$_1$, AF$_2$ or U$_2$), and zero if they are in any of the other two.
In each Monte-Carlo step, the order parameter is calculated for every unit cell.
It should be noted that when averaging  these parameters for different copies we will take the absolute value of the MC measurement, since otherwise it will average to zero.

\subsubsection{Wang-Landau}
\label{sec:MC-WL}

The Wang-Landau (MC-WL) algorithm [\onlinecite{WangLandau01}] has emerged as an efficient Monte-Carlo technique in
statistical physics. In the last years, this technique has been applied to a variety of 
studies of classical statistical models such as the Ising [\onlinecite{IsingWL2002}] and Potts [\onlinecite{Potts2001}] spin model,
Heisenberg ferromagnetic systems [\onlinecite{FM2006}] and
antiferromagnetic frustrated models [\onlinecite{SpinIceFerreyra}]. 
In this work, we performed simulations on lattices of $N=4\times L^2$ sites ($L=2-10$). To optimize the convergence 
of the algorithm we use the modification proposed in Ref.[\onlinecite{Belardinelli07}].
This algorithm allows  the estimation of the energy density of states (DoS) $g(E)$ performing a random walk in energy space.
Then, from the DoS we can construct the partition function and obtain thermodynamic
quantities like 
entropy and free energy, which are not easily accessible through conventional Monte-Carlo methods based on Metropolis
algorithm. To better characterize the system, we often  need to calculate a joint density of states (JDoS) $g(E,\text{OP})$, where $\text{OP}$ is an
order parameter. This allows us to explore the phases of the system and also to calculate thermodynamic quantities like the Landau free energy[\onlinecite{LandauFreeE}].

Once obtained $g(E,\text{OP})$,  the partition function can be computed as 
\begin{equation}
Z(\beta,\mu)=\sum_{E,\text{OP}}g(E,\text{OP})\,e^{-\beta(E-\,\mu\,\text{OP})}
\end{equation}

\noindent with $\beta=1/k_BT$ and $\mu$ some Lagrange multiplier (for example,
$\mu$ may be the magnetic field and thus the  $\text{OP}$ would correspond to the total magnetisation) [\onlinecite{foot2}].
From the partition function, we can obtain thermodynamic quantities in the canonical ensemble for all values of $\beta$ (temperature) and $\mu$. 
For instance, the mean value of the energy $E$ and the order parameter $OP$ may be calculated as:
\begin{eqnarray}
\langle E\rangle&=&\frac{1}{Z(\beta,\mu)}\sum_{E,\text{OP}}g(E,\text{OP})E\,e^{-\beta(E-\,\mu\,\text{OP})}\\
\langle \text{OP}\rangle&=&\frac{1}{Z(\beta,\mu)}\sum_{E,\text{OP}}g(E,\text{OP})\text{OP}\,e^{-\beta(E-\,\mu\,\text{OP})}
\end{eqnarray}
In addition to the standard averages, it is straightforward to determine some important quantities 
like Helmholtz's free energy and entropy defined as

\begin{eqnarray}
F(\beta,\mu) &=&-\beta^{-1}\ln(Z(\beta,\mu))\\
S(\beta,\mu) &=&\beta(\langle E\rangle-F(\beta,\mu)).
\end{eqnarray}

Finally, to obtain more information on the global behavior of the order parameter around the phase transition, in this work we have calculated
the Landau free energy as

\begin{eqnarray}
e^{-\beta\,F_{L}(\beta,\text{OP})}&=&\sum_{E}g(E,\text{OP})e^{-\beta\, E}
\label{eq:LandauFreeE}
\end{eqnarray}
The study of the Landau free energy will be further discussed in the following section.

\section{Results and phase diagrams}
\label{sec:results}
In this section we  explore the low temperature behavior of the model  introduced in Eq. (\ref{eq:Hamiltonian})
combining the three approaches described in the previous section.  
We first focus on the model in the absence of a magnetic field, where a rich phenomenology is found. Then, we discuss the effect of an external field.

\subsection{$h=0$}
\label{sec:Results_h0}

In the absence of the magnetic field, the system presents different phase transitions and selection mechanisms, which we will discuss below. 
\subsubsection{ $J_1\neq J_x$: second order phase transitions for broken $\mathcal{Z}_2$ symmetry ground states }
\begin{figure}[h!]
\includegraphics[width=1\columnwidth]{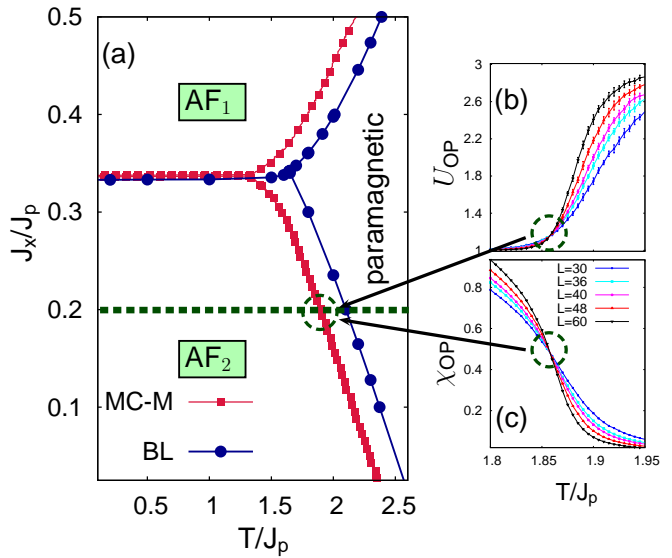}
\caption{\label{fig:J1JxVsT}(Color online) (a): $J_x/J_p$ vs $T/J_p$ phase diagram, for $J_1=J_p$ for Bethe lattice calculations (full (blue) circles)
and Metropolis  simulations (full (red) squares) for $L=48$. 
The transition lines between the paramagnetic and the ordered phases are of second order as it is confirmed by the Binder cumulant 
(b) and scaled order parameter susceptibility (c) for  $(J_x/J_p=0.2,J_1/J_p=1)$.}
\end{figure}

 As a typical example, we focus 
on the $J_1=J_p$ case, where for $J_x<J_p/3$ the ground state is AF$_2$ and for $J_x>J_p/3$, AF$_1$. 
 We show the transition lines from a paramagnetic to an ordered phase in a $J_x/J_p$
vs $T/J_p$ phase diagram in Fig.\ref{fig:J1JxVsT} [\onlinecite{foot3}]. 
These lines were obtained from the Bethe lattice analysis 
and from MC-M simulations. According to the Bethe lattice technique, these transitions are of second order. 
We check this with MC-M following the standard procedure: locating the  crossing point of the
corresponding susceptibility $\chi_{\text{OP}}$ and Binder cumulant $U_{\text{OP}}$  (measured for different system sizes) for the relevant order parameters, 
defined as:

\begin{equation}\label{eq:Binder}
 \chi_{\text{OP}}=\frac{N}{T}\langle \text{OP}^2\rangle\,\,\,\,U_{\text{OP}}=\frac{\langle \text{OP}^4\rangle}{(\langle \text{OP}^2\rangle)^2}
\end{equation}
 These phase transitions are associated with the breaking of  $\mathcal{Z}_2$ symmetry. Therefore, the critical exponent for the susceptibility 
near the critical temperature is known,
$\eta=\frac{1}{4}$[\onlinecite{Pathria}]. In that region, $\chi_{\text{OP}} = L^{2-\eta}(f(|1-\frac{T}{T_c}|L^{1/\nu}))$. 
Thus the scaled susceptibility is size independent at the critical temperature $T_c$, where  different system sizes should show a crossing  point.
We illustrate this method for the AF$_2$ in  Fig. \ref{fig:J1JxVsT} (b and c)  where we  show the  Binder cumulant (b)
and the scaled susceptibility (c) for different system sizes as a function of $T/J_p$  ($J_x/J_p=0.2$,$J_1/J_p=1$). 
We can observe that indeed the curves for different $L$ plotted as functions of $T/J_p$ for both the Binder cumulant and the normalized susceptibility 
exhibit a crossing point at the critical temperature, confirming that the transition is of second order. 
Finally, we remark that the agreement between the Bethe lattice results and the MC-M simulations is both qualitative and quantitative, as
can be seen in Fig. \ref{fig:J1JxVsT}: the difference in the values of the critical temperatures is $~10\%$.  This difference is probably
due to both the finite size effect in the MC-M simulations and the fact that the BL analysis is a mean field calculation.

\subsubsection{Strong frustrated line $J_1=J_x$: first and second order phase transitions, cooperative paramagnet phase}

\begin{figure}[h!] 
\includegraphics[width=0.9\columnwidth]{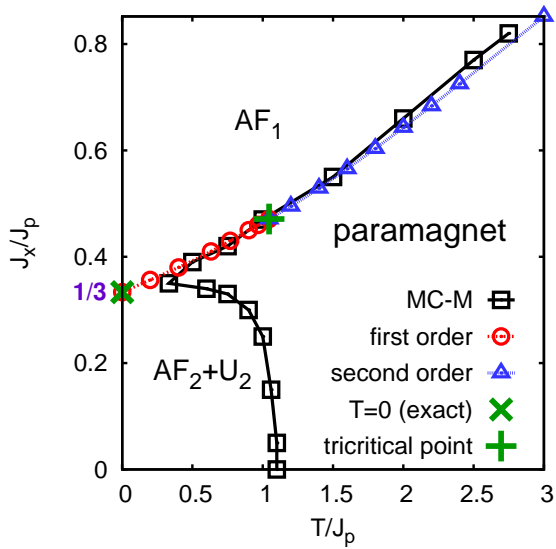}
\caption{\label{fig:PD_J1JxT}(Color online) Phase diagram  obtained by BL and MC-M methods for $h = 0$ and $J_1 = J_x$. 
The empty (blue) triangles joined by a finely dashed line represent a second order phase transition and the empty (red) circles joined by a dashed line
 represent first order transitions. The $+$ (green) symbol is a tricritical point and the $\times$ (green) symbol corresponds
 to the exact $T = 0$ first order point. 
 The empty (black) squares correspond to the maximum of the specific heat  for $L=48$  MC-M simulations} 
\end{figure}

We now focus on the $J_1=J_x$ line, where the system exhibits a range of interesting and different phenomena at low temperatures. Our main results 
are summarized in the $J_1/J_p=J_x/J_p$ vs $T/J_p$ phase diagram in Fig. \ref{fig:PD_J1JxT}. The transition lines and the tricritical point were
obtained with the Bethe lattice technique.
The points   correspond to the maximum of the specific heat in the MC-M simulations for $L=48$. We can identify three types
of behavior: (i) a cooperative paramagnet phase 
for $J_1=J_x < J_p/3$,
(ii) a first order phase transition from the paramagnetic phase to the AF$_1$ phase for $J_p/3 < J_1=J_x < J^* \approx 0.45J_p$, 
and (iii) a second order phase transition from a paramagnetic to a broken $\mathcal{Z}_2$ symmetry phase (AF$_1$) for $J_1=J_x > J^*$.
The highly frustrated point $J_1=J_x=J_p/3$ will be discussed in the next subsection.

To further characterize these three types of behaviour, we study  several variables.
Besides the OP$_{\text{AF}_1}$ parameter, the specific heat and the entropy,
we introduce a variable that we call ``$J_p$ correlator'' (CJ$_p$) defined in terms of the scalar product of spins connected by $J_p$ as,
\begin{equation}
 \text{CJ}_p=\frac{1}{2} \sum_{\vr}\langle \sigma^A_{\vr}\sigma^C_{\vr} + \sigma^B_{\vr}\sigma^D_{\vr} \rangle
 \label{eq:CJcorrelator}
\end{equation}
\noindent This variable is defined so that CJ$_p=-1$ for both AF$_2$ and U$_2$, and CJ$_p=+1$ in AF$_1$. This will be relevant to identify the cooperative
paramagnet phase, as we discuss below. 

\begin{figure*}[t!]
\includegraphics[width=1.8\columnwidth]{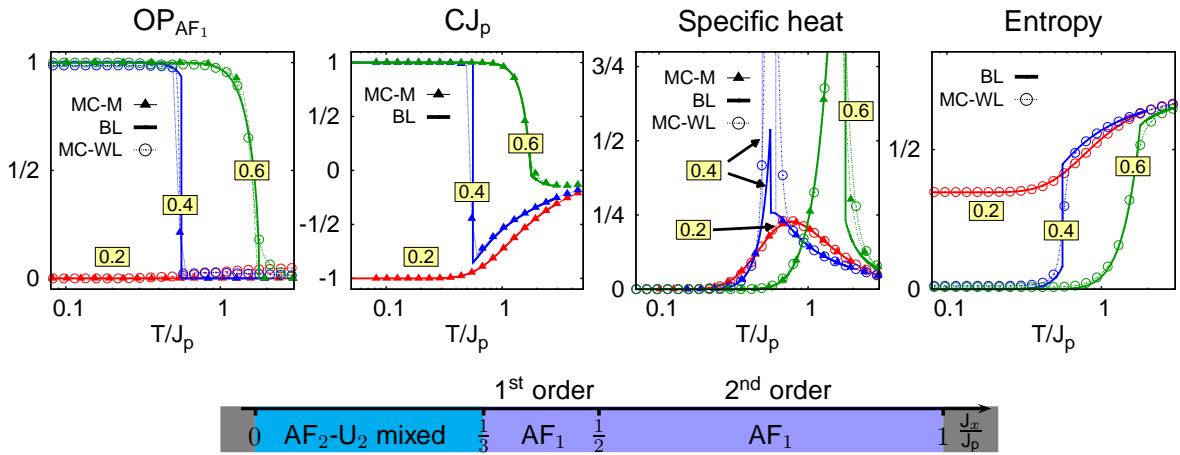}
\caption{\label{fig:J1Jxline} (Color online) Curves of AF$_1$ order parameter, CJ$_p$ correlator,  
specific heat  and entropy (in units of $\ln 2/N$) as a function of $T/J_p$  for three values of the couplings (indicated in each set of curves)
along the $J_1=J_x$ line:
$J_x=0.2\,J_p<J_p/3$ (red), $J_p/3<J_x=0.4\,J_p<J^*$ (blue) and $J^*<J_x=0.6\,J_p$  (green) obtained by comparison of Metropolis simulations (MC-M)
(full triangles), Bethe lattice approximation (BL) (full line) and Wang-Landau (MC-WL) (empty circles).   
In each region a  different behavior is seen. For the larger values of $J_x/J_p$ the system orders 
in the AF$_1$ phase through a second order transition from the paramagnetic phase. For a smaller range of $J_1=J_x>J_p/3$, 
the system orders in the AF$_1$ at low temperatures through a first order transition, clearly seen as a jump in the order parameter 
and an off-scale peak in the specific heat. For $J_1=J_x<J_p/3$, the system does not order at low temperature remaining in a macroscopically degenerate state. 
 In the latter case, the specific heat shows a broad peak, as seen for example in spin ice systems.   The entropy in this region does
not vanish as the temperature is lowered towards $T = 0$, but  tends to the residual value $s=1/2$.}

\end{figure*}

 We now comment specifically on the physics of each regime of the couplings.  
To illustrate each case, we show for a specific point in each range of the couplings ($J_1=J_x=0.2J_p,0.4J_p,0.6J_p$) 
the  OP$_{\text{AF}_1}$ parameter, the specific heat, the entropy and  CJ$_p$
as a function of temperature in Fig.\ref{fig:J1Jxline}.

\begin{itemize}
\item[(i)] For $J_1=J_x<J_p/3$,  at low temperatures, the system is in a cooperative paramagnet phase.
 It is degenerate: 
the plaquettes are in a mixture of AF$_2$ and U$_2$ states.
A first indicator of this phase is the behavior of the order parameters and the CJ$_p$ correlator at low temperatures.
 $\text{OP}_{\text{AF}_1},\text{OP}_{\text{AF}_2},\text{OP}_{\text{U}_2}$
average up to 0, but the CJ$_p$ correlator tends to $-1$. This indicates that the pairs of spins joined by $J_p$ are either $+-$ or $-+$, and thus that
each plaquette is either in an AF$_2$ or a U$_2$ state.
Another important indicator of the cooperative paramagnet phase is the shape of the specific heat. 
It can be seen that in this case it shows a broad maximum, there is no sharp feature. This kind of behavior is seen for example in spin ice systems 
[\onlinecite{Diep,Bramwell01,Harris97,Bramwell01_2,Ramirez99,Fennell02,Zvyagin}]:
it is an indication that no long range order is developed through a thermodynamic phase transition.
This cooperative paramagnet phase has extensive entropy. 
We show the entropy as a function of temperature, obtained from Bethe lattice and MC-WL calculations (Fig.\ref{fig:J1Jxline}, third panel from the left). 
All the curves have the same $T\to\infty$ limit ($1$ in units of $N\log 2$) as is expected. However, at low temperatures, 
the red curve shows that for $J_1=J_x<J_p/3$ the system remains disordered as a consequence of plaquette degeneration,
and that at low temperatures $s=\frac{1}{2}$.
\item[(ii)] For $J_p/3<J_1=J_x<J^*$, there is a sharp first order transition to the ordered phase (AF$_1$) at $T/J_p \sim 0.55$. 
This is clearly seen as a characteristic jump in the parameters shown in Fig. \ref{fig:J1Jxline} (blue lines).
The specific heat obtained from both the
MC-M and MC-WL simulations shows a clear discontinuity (the peak in the simulations is off-scale, and therefore not shown in the figure). 
\item[(iii)] For $J_1=J_x>J^*$, the system is also ordered at low temperatures: all the plaquettes are in the AF$_1$ case. The transition from the paramagnetic 
to the ordered phase  is of second order. We have confirmed this by computation of the scaled susceptibility and the Binder cumulant
for different system sizes, as done in the previous subsection. 
\end{itemize}

\begin{figure}[h!]
\includegraphics[width=0.95\columnwidth]{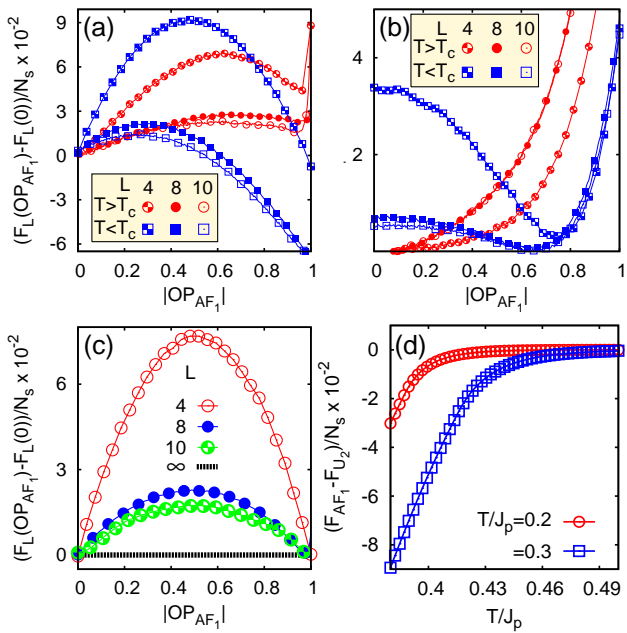}
\caption{\label{fig:FreeEnergy} (Color online) First three panels (a-c): MC-WL simulation results 
for free energy vs order parameter OP$_{\text{AF}_1}$ around the critical temperature $T_c$. $T>T_c$ is indicated by circles and $T<T_c$ by squares, 
for different system sizes $L$.
(a) for $J_1=J_x=0.4J_p$,  where there is a first order transition; (b) $J_1=J_x=0.6J_p$,   where there is a second order transition. 
(c) Landau free energy at the critical temperature for different sizes for a first order phase transition. As the size 
of the system is increased, the Landau free energy is flatter. The dotted flat line is the behavior in the thermodynamic limit.
Fourth panel (d) difference of the free energy of the AF$_1$ and the U$_2$ states as a function of the $J_x$ parameter along the degenerate
line $J_1=J_p/3; J_x>J_p/3$ for two different temperatures obtained from the Bethe lattice approximation.}
\end{figure}

We now exploit the power of the MC-WL technique, studying the Landau free-energy (Eq. (\ref{eq:LandauFreeE})) as a function of
the relevant order parameter. We will discuss how using this variable it is possible to provide further evidence of the different types
of phase transitions for $J_1=J_x>J_p/3$.   The Landau free-energy as a function of OP$_{\text{AF}_1}$, for two values of the 
couplings characteristic of each region,  is shown in Fig. \ref{fig:FreeEnergy}.
On one hand, in  Fig. \ref{fig:FreeEnergy} (a) we see that for $J_x/J_p=0.4$ the behavior near the critical temperature $T_c$ 
the position of the global minimum changes abruptly from $\langle \text{OP}_{\text{AF}_1}\rangle_{\text{min}}=0$, for $T>T_c$,
to $\langle \text{OP}_{\text{AF}_1}\rangle_{\text{min}}=1$ (normalized) for $T<T_c$. 
At this point it should be clarified that this shape of the Landau free-energy is characteristic of a finite system size 
(see for example Ref. [\onlinecite{LFE_comment}]), as can be seen in   Fig. \ref{fig:FreeEnergy}  (c), where the Landau free energy
for different system sizes is shown. The curves are flatter with increasing system size $L$.

In the thermodynamic limit, the curve at  $T=T_c$ is expected to tend  to the dotted curve which has a flat portion  between the two minima. 
On the other hand, typical second-order transition behavior is observed for  $J_x/J_p=0.6$ (Fig.  \ref{fig:FreeEnergy} (c)) with a 
gradual increase of $\langle \text{OP}_{\text{AF}_1}\rangle_{\text{min}}$ from zero, 
for $T>T_c$ to $\langle \text{OP}_{\text{AF}_1}\rangle_{\text{min}}\approx 0$ for $T\lesssim T_c$. 
This analysis supports previous results obtained with BL and MC-M, and  provides a different way to study the nature of the phase transition.
\subsubsection{Highly frustrated point $J_1=J_x=J_p/3$: partial order-by-disorder}

 Let us focus on the highly frustrated point,  $J_1=J_x=J_p/3$. At this point, 
the three $m_0=0$ configurations (AF$_1$, AF$_2$ and U$_2$) 
in each plaquette have the same energy. However, there is a substantial difference between the possible configurations.
In the AF$_1$ case, all plaquettes are in the same arrangement,
the degeneracy is only two-fold. In the AF$_2$ and U$_2$ case, as was discussed for  $J_1=J_x<J_p/3$, the plaquettes can be in any of the two
arrangements, thus leading to extensively degenerate ground states.
In order to clarify if the system chooses one of these ground states with temperature  
(and thus thermal order by disorder [\onlinecite{Villain89,Shender82}] is at play), 
we study the three order parameters (Eqs. (\ref{eq:OP1}), (\ref{eq:OP2}), (\ref{eq:OP3})), 
the CJ$_p$ correlator (Eq. (\ref{eq:CJcorrelator})), the specific heat and the entropy as a function of temperature at $h=0$. 
The results are analogous to those in Fig. \ref{fig:J1Jxline} (dotted red line) for the case $J_x/J_p<1/3$. 
 This indicates a phenomenon that is not present in the previously discussed case: (partial) thermal order-by-disorder.
At low temperatures the system chooses the  cooperative paramagnet AF$_2$-U$_2$ phase over the AF$_1$ one,
since the AF$_2$-U$_2$ phase has a lower contribution to the free energy.

\subsubsection{Degenerate line $J_{1(x)}=J_p/3, J_{x(1)}>J_p/3$: order-by-disorder}

Now, we  center our study in the line $J_{1(x)}=J_p/3, J_{x(1)}>J_p/3$ (horizontal and vertical lines in Fig. \ref{fig:phdiagT0}), 
where the ground state of the system is either in a AF$_1$ state or  a U$_2$(AF$_2$) state. 
Since the two cases are equivalent, we direct our attention to the $J_1=J_p/3, J_x > J_p/3$ case.
Contrary to the highly frustrated point, in this case both phases have entropy $s=0$ in the thermodynamic limit.
However, MC-M simulations show that for lower values of $J_x$ the system choses the U$_2$ phase, whereas for higher values
it can be in either state. This is evidence of order-by-disorder state selection at low $J_x$. The origin of this selection can be understood in the following way.
In the order-by-disorder phenomenon, in the $T\to0$ limit, the system chooses states which have a lower contribution to the free energy,
even though they are degenerate at $T=0$. In this case, the U$_2$ state has lower energy fluctuations,
and thus it has a lower free energy at $T\to0$. For both U$_2$ and AF$_1$  states,
the first excitation is simply to flip one spin, and in both cases 
the energy change is $6J_x$. To explore the next excitations to the ground states, one can consider the energy difference of flipping spins joined by
the different couplings. For the AF$_1$ state, the next lowest energy excitation is flipping along a $J_x$ bond with an energy cost of $\Delta E^{(J_x)}=8J_x$.
For the U$_2$ case for low values of $J_x$ flipping along a $J_p$ bond has a lower energy cost $\Delta E^{(J_p)}=12(J_x-J_p/3)$.
Therefore, when the temperature is low enough to make these excitations available, the systems chooses the U$_2$ state.
The Bethe lattice technique provides an interesting way of checking this, since the free energy of each state can be calculated.
Fig. \ref{fig:FreeEnergy} (d)
shows the difference in the free energy between the AF$_1$ state and the U$_2$ state as a function of $J_x$ for different temperatures.
It can be clearly seen that the U$_2$  state has a lower free energy, but that this
difference is smaller at lower temperatures with increasing $J_x$.

\subsection{$h>0$}

\begin{figure}[h!] 
 \includegraphics[width=1\columnwidth]{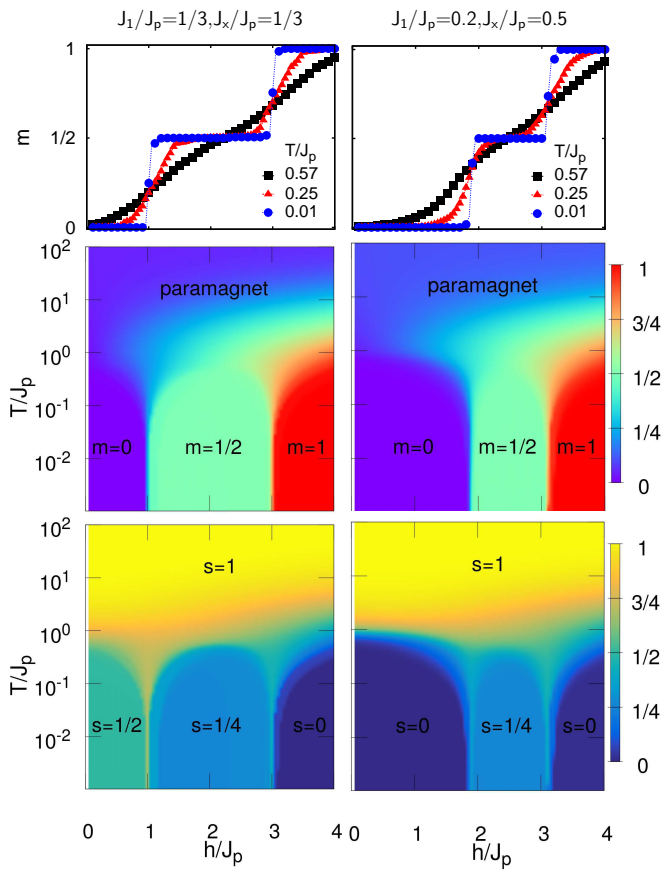}
\caption{\label{fig:J1JxTercio_hvsT}(Color online)  Magnetisation $m$ and entropy  per spin (in units of $\ln(2)$) $s$ 
as a function of the magnetic field $h/J_p$ obtained with Wang-Landau simulations 
for the highly frustrated point $J_1=J_x=J_p/3$ (left) and for an arbitrary value of the couplings $J_1=0.2J_p, J_x=0.5J_p$ (right).
The top panel shows the magnetisation curves at different temperatures. The middle and bottom panels are
the density plots corresponding  to  $h/J_p$ vs $T/J_p$ phase diagrams for $m$ and $s$, respectively.
}
\end{figure}

In previous sections we have studied the effect of the thermal fluctuations in the stability and transitions to the low temperature phases.
The coupling with an external magnetic field can tune the system into high energy plaquette phases inaccessible at $h=0$.
Here  we will compare a typical case with  a well defined $m=0$ state, for example $J_1=0.2J_p, J_x=0.5J_p$
with the highly frustrated case $J_1=J_x=J_p/3$.
 The magnetisation and the entropy as a function of temperature and magnetic field by MC-WL simulations are shown
in Fig.  \ref{fig:J1JxTercio_hvsT}. The $m=0$ phases follow the physics previously discussed for $h=0$.
The transition from the paramagnetic to the $m=1/2$ phase is predicted to be of second-order according to the Bethe lattice calculations,
which we confirmed with MC-M simulations using the susceptibility and Binder-cumulant analysis as before. A remarkable feature is that 
 the $m=1/2$ plateau has finite extensive entropy
for both the highly frustrated case and the more typical case presented here, 
as commented in Sec.\ref{sec:PhaseDiagramT0}.
Finally, high magnetic fields stabilize the saturation plateau $m=1$. 
\section{Conclusions}
\label{sec:Conclusions}
In this work, we present a complete study of the  Ising model on a bilayer honeycomb lattice  
including interlayer frustration and coupling with an external magnetic field. We first present and discuss the exact $T=0$ phase diagram and we highlight
specific points and lines where we expect interesting physics.
Then, we study the effect of temperature using a combination of analytical mean-field-like considerations (Bethe lattice) and  Monte-Carlo (Metropolis and Wang-Landau) simulations.
The interplay between these techniques has been essential to obtain 
 magnetic and thermodynamic properties of the system.
We have found a very rich phase diagram with nontrivial regions characterized by broken symmetries and non zero entropy values. 

In the case of zero magnetic field, along the highly frustrated line where the intralayer ($J_1$) and frustrating interlayer ($J_x$) couplings are equal, for $J_1=J_x<J_p/3$
there is a crossover from a paramagnetic phase
to a cooperative paramagnet, where there is finite entropy. At the highly frustrated point  $J_1=J_x=J_p/3$, order-by-disorder is at play,
and this cooperative paramagnet phase is the one selected at low temperatures. For higher values of the couplings, the system is ordered at low
temperatures. The Bethe lattice technique shows that this transition is of first order for a certain range of parameters, and then it is second order.
We checked this studying thermodynamic variables with both types of simulations. We also used the Wang-Landau technique to study the Landau-free-energy with the 
corresponding order parameter, to illustrate the difference between a first and a second order phase transitions.
Through Monte-Carlo Metropolis simulations and the free energy obtained from the Bethe lattice approximation we showed that order by disorder is also at play
in the coexistence lines between two ordered phases ($J_{1(x)}=J_p/3,J_{x(1)}>J_p/3$).

In the presence of a magnetic field, there are three plateaux: at zero and $1/2$ magnetisation, and the saturation plateau. The $1/2$ plateau
is highly degenerate for all antiferromagnetic couplings. This implies that there is 
a non zero entropy induced by the field through a second order
phase transition, even for sets of
parameters where in the absence of the external field the system is ordered.

In summary,  we have shown that the analytical (Bethe lattice) and numerical (Metropolis and Wang-Landau simulations)
techniques are complementary and provide a solid way
of exploring the different non trivial phases of this system.
We expect to extend this sort of study to other highly degenerate systems, like the highly frustrated honeycomb
and kagome lattices in the extended Heisenberg model, exploiting the richness of these techniques
to obtain the complete phase diagrams

\section{Acknowledgements}

The authors thank Sergio Cannas, Rodolfo Borzi and Mar\'ia Victoria Ferreyra for fruitful discussions. F.A.G.A. and H.D.R. are partially supported by CONICET (PIP 2015-0813), 
ANPCyT (PICT 2012-1724) and SECyT-UNLP. P.S. is partially supported by CONICET (PIP 11220150100327) and SECyT-UNC.

\bigskip

\appendix
\section{Matrix Elements for Spherical approximation}
\label{AppendixA}

The explicit form of the   the analytical expressions for the boundaries for the plateaux at $T=0$ (Fig. \ref{fig:phdiagT0}) are given 
in the table below, where $\alpha=J_x/J_1$

\begin{table}[h!]
\centering
\begin{tabular}{ |c|c|c|c| } 
\hline
&\multicolumn{3}{|c|}{critical field $h_c$}\\
case & AF$_2$-UAF & AF$_1$-UAF& UAF-U \\
\hline
(b) $J_x=J_1$ & $1$* & $-1+6\frac{J_1}{J_p}$ &$1+6\frac{J_1}{J_p}$ \\ 
(c) $J_x < J_1$& $-1 + 3(\alpha+1)\frac{J_1}{J_p}$ & $1+3(1-\alpha)\frac{J_1}{J_p}$ &$1+3(1+\alpha)\frac{J_1}{J_p}$\\ 
(d) $J_x > J_1$& $-1 + 3(\alpha+1)\frac{J_1}{J_p}$ & $1+3(\alpha-1)\frac{J_1}{J_p}$ & $1+3(1+\alpha)\frac{J_1}{J_p}$\\ 
\hline
\end{tabular}
\caption{(*) in this case the $h_c$ corresponds to the intersection of U$_2$/AF$_2$-UAF phases.}
\end{table}
\vspace{2cm}

\section{Bethe lattice calculations}
\label{Appendix B}
%
The Bethe lattice allows us to write down recursion relations (RR) for the  eight 
partial partition functions (PPF)

\begin{widetext}
\begin{subequations} \label{errz}
\begin{eqnarray}
Z^{a'}_{++}   &=& e^{-k_p-2 k_1-2 k_x+2 h} (Z^{b}_{++})^2 + e^{k_p}  
((Z^{b}_{+-})^2+(Z^{b}_{-+})^2)+ e^{-k_p+2 k_1+2 k_x-2 h} (Z^{b}_{--})^2 \\
Z^{a'}_{+-}  &=& e^{-k_p+2 h} (Z^{b}_{++})^2 + e^{k_p-2 k_1+2k_x}  (Z^{b}_{+-})^2+
 e^{k_p+2 k_1-2 k_x} (Z^{b}_{-+})^2+
 e^{-k_p-2 h} (Z^{b}_{--})^2 \\
Z^{a'}_{-+}   &=& e^{-k_p+2 h} (Z^{b}_{++})^2 +  e^{k_p+2 k_1-2 k_x}(Z^{b}_{+-})^2+
e^{k_p-2 k_1+2k_x} (Z^{b}_{-+})^2+
 e^{-k_p-2 h} (Z^{b}_{--})^2 \\
Z^{a'}_{--}   &=& e^{-k_p+2 k_1+2 k_x+2 h} (Z^{b}_{++})^2 + e^{k_p}  ((Z^{b}_{+-})^2+(Z^{b}_{-+})^2)+
 e^{-k_p-2 k_1-2 k_x-2 h} (Z^{b}_{--})^2 \\
Z^{b'}_{++}   &=& e^{-k_p-2 k_1-2 k_x+2 h} (Z^{a}_{++})^2 + e^{k_p}  ((Z^{a}_{+-})^2+(Z^{a}_{-+})^2)+
 e^{-k_p+2 k_1+2 k_x-2 h} (Z^{a}_{--})^2 \\
Z^{b'}_{+-}   &=& e^{-k_p+2 h} (Z^{a}_{++})^2 + e^{k_p-2 k_1+2k_x}  (Z^{a}_{+-})^2+
 e^{k_p+2 k_1-2 k_x} (Z^{a}_{-+})^2+
 e^{-k_p-2 h} (Z^{a}_{--})^2 \\
Z^{b'}_{-+}   &=& e^{-k_p+2 h} (Z^{a}_{++})^2 +  e^{k_p+2 k_1-2 k_x}(Z^{a}_{+-})^2+
e^{k_p-2 k_1+2k_x} (Z^{a}_{-+})^2+
 e^{-k_p-2 h} (Z^{a}_{--})^2 \\
Z^{b'}_{--}   &=& e^{-k_p+2 k_1+2 k_x+2 h} (Z^{a}_{++})^2 + e^{k_p}  ((Z^{a}_{+-})^2+(Z^{a}_{-+})^2)+
 e^{-k_p-2 k_1-2 k_x-2 h} (Z^{a}_{--})^2 \,.
\end{eqnarray}
\end{subequations}
\end{widetext}

\noindent The RR  Eqs. ({\ref{errz}) are divergent, and, as usual, 
we proceed to define new recursion relations that converge in the thermodynamic limit
  dividing by $Z^{a}_{--}$ or $Z^{b}_{--}$,
  
\newpage

\begin{widetext}
\begin{equation}\label{edr}
R_1\,=\,\frac{Z^{a}_{++}}{Z^{a}_{--}}\;;\;
R_2\,=\,\frac{Z^{a}_{+-}}{Z^{a}_{--}}\;;\;R_3\,=\,\frac{Z^{a}_{-+}}{Z^{a}_{--}}\;;\; R_4\,=\,\frac{Z^{b}_{++}}{Z^{b}_{--}}\;;\;
R_5\,=\,\frac{Z^{b}_{+-}}{Z^{b}_{--}}\;;\;R_6\,=\,\frac{Z^{b}_{-+}}{Z^{b}_{--}}
\end{equation}
\end{widetext}

Eqs. (\ref{errz}) and (\ref{edr}) give the following RR

\begin{widetext}
\begin{subequations} \label{errr}
\begin{eqnarray}
R'_{1}&=&\frac{1}{D_{b}} \left(e^{-k_p-2 k_1-2 k_x+2 h} R_4^2 + e^{k_p} (R_5^2+R_6^2)+
 e^{-k_p+2 k_1+2 k_x-2 h}\right) \\
R'_{2}&=&\frac{1}{D_{b}} \left(e^{-k_p+2 h} R_4^2 + e^{k_p-2 k_1+2k_x}  R_5^2+
 e^{k_p+2 k_1-2 k_x} R_6^2+
 e^{-k_p-2 h} \right) \\
R'_{3}  &=& \frac{1}{D_{b}} \left( e^{-k_p+2 h} R_4^2 +  e^{k_p+2 k_1-2 k_x}R_5^2+
e^{k_p-2 k_1+2k_x} R_6^2+
 e^{-k_p-2 h} \right) \\
R'_{4}&=&\frac{1}{D_{a}} \left(e^{-k_p-2 k_1-2 k_x+2 h} R_1^2 + e^{k_p} (R_2^2+R_3^2)+
 e^{-k_p+2 k_1+2 k_x-2 h}\right) \\
R'_{5}&=&\frac{1}{D_{a}} \left(e^{-k_p+2 h} R_1^2 + e^{k_p-2 k_1+2k_x}  R_2^2+
 e^{k_p+2 k_1-2 k_x} R_3^2+
 e^{-k_p-2 h} \right) \\
R'_{6}  &=& \frac{1}{D_{a}} \left( e^{-k_p+2 h} R_1^2 +  e^{k_p+2 k_1-2 k_x}R_2^2+
e^{k_p-2 k_1+2k_x} R_3^2+
 e^{-k_p-2 h} \right) \,,
\end{eqnarray}
\end{subequations}
\end{widetext}

\noindent where

\begin{widetext}
\begin{subequations} \label{ed}
\begin{eqnarray}
D_{a}&=&e^{-k_p+2 k_1+2 k_x+2 h} R_1^2 + e^{k_p}  (R_2^2+R_3^2)+
 e^{-k_p-2 k_1-2 k_x-2 h}\\
D_{b}&=&e^{-k_p+2 k_1+2 k_x+2 h} R_4^2 + e^{k_p}  (R_5^2+R_6^2)+
 e^{-k_p-2 k_1-2 k_x-2 h}\,.
\end{eqnarray}
\end{subequations}
\end{widetext}

\subsection{ Fixed points and the thermodynamic phases}

The thermodynamic phases are given by the stable fixed points $\vec{R}^*$ of
Eqs.(\ref{errr}). The continuous, or second order lines are defined as coincident
stability lines of different fixed points (phases).

At zero magnetic field, the fixed points of the different phases are given by
the conditions

\begin{itemize}
\item Paramagnetic phase 
\begin{equation} \label{epp}
R_1^*=1 \;;\; R_3^*=R_2^*\;;\; R_4^*=1 \;;\;
R_5^*=R_2^*\;;\;R_6^*=R_2^*
\end{equation}

\item $F_2$ phase 
\begin{equation} \label{ef2p}
R_1^*=1 \;;\; R_5^*=R_2^*\;;\; R_4^*=1 \;;\;R_6^*=R_3^*
\end{equation}

\item $AF_1$ phase
\begin{equation} \label{eaf1p}
R_1^*\ne 1 \;;\; R_3^*=R_2^*\ne R_5^*=R_6^*\;;\; R_4^*\ne1
\;;\; R_4^*\ne R_1^* 
\end{equation}

\item $AF_2$ phase.
\begin{equation} \label{eaf2p}
R_1^*=1 \;;\; R_6^*=R_2^*\;;\; R_4^*=1 \;;\;R_5^*=R_3^*
\end{equation}

\end{itemize}

\noindent For $H\ne 0$ the symmetry is broken, and these relations between the 
$R_i^*$ are not valid.

\subsection{ The partition function, thermodynamic averages and the free energy}

In order to classify the different thermodynamics phases, we need the partition
function  and thermodynamics averages, as the magnetisation per site. 
When  the stability lines of two 
(or more) fixed points are not coincident,  the overlap region is a coexistence zone,
 and we also need to calculate the first-order line as the line where the free 
energies   of the corresponding phases take the same value.

As usual, the thermodynamics averages and the free energy must be calculated
on the central region. There is not a unique manner to define the central zone,
we define it as a central plaquette where  four subtrees are attached as it shown
as shown in (Fig. 3).

Putting a plaquette as central zone, always a bond belong to
the sublattice $a$ ($b$), and we attach to them two subtrees with the root
belonging to the sublattice $b$ ($a$), then, at any generations, including the
surface, a half of points belong to a sublattice and the other half to the
other sublattice. Then, we obtain for the partition function for a $M-$generations
tree,

\begin{widetext}
\begin{eqnarray}  \label{ezp}
{\cal Z}_M&=&(Z^{a}_{++})^2 \left(e^{-2 k_p-2 k_1-2 k_x+4 h} (Z^{b}_{++})^2 +
e^{2 h} \left( (Z^{b}_{+-})^2 +(Z^{b}_{-+})^2 \right)+e^{-2 k_p+2 k_1+2 k_x}
(Z^{b}_{--})^2 \right) \nonumber \\
\mbox{}&+&(Z^{a}_{+-})^2 \left(e^{2 h} (Z^{b}_{++})^2 +
e^{2 k_p-2 k_1+2k_x} (Z^{b}_{+-})^2 + e^{2 k_p+2 k_1-2 k_x}(Z^{b}_{-+})^2 +
e^{-2 h} (Z^{b}_{--})^2 \right) \nonumber  \\
\mbox{}&+&(Z^{a}_{-+})^2 \left(e^{2 h} (Z^{b}_{++})^2 +
e^{2 k_p+2 k_1-2k_x} (Z^{b}_{+-})^2 + e^{2 k_p-2 k_1+2 k_x}(Z^{b}_{-+})^2 +
e^{-2 h} (Z^{b}_{--})^2 \right)   \\
\mbox{}&+&(Z^{a}_{--})^2 \left(e^{-2 k_p+2 k_1+2 k_x} (Z^{b}_{++})^2 +
e^{-2 h} \left( (Z^{b}_{+-})^2 +(Z^{b}_{-+})^2 \right)+e^{-2 k_p-2 k_1-2 k_x-4 h}
(Z^{b}_{--})^2 \right)\,. \nonumber
\end{eqnarray}
\end{widetext}

\noindent where all the PPF corresponds to M-generations subtrees.

The magnetisations in the four different sites 
$m_i=\langle \sigma_i\rangle$, where  $i=A,B,C,D$ number the four sites of the
central plaquette, take now  the expressions

\begin{widetext}
\begin{subequations} 
\begin{eqnarray}  \label{emp}
m_A\,=\,\frac{1}{\cal Y} &&\left\{ 
R_1^{*2} \left(e^{-2 k_p-2 k_1-2 k_x+4 h} R_4^{*2} +
e^{2 h} \left( R_5^{*2} +R_6^{*2} \right)+e^{-2 k_p+2 k_1+2 k_x}
\right) \right.\nonumber \\
\mbox{}&+&R_2^{*2}  \left(e^{2 h} R_4^{*2} +
e^{2 k_p-2 k_1+2k_x} R_5^{*2} + e^{2 k_p+2 k_1-2 k_x} R_6^{*2} +e^{-2 h}  \right) 
\nonumber  \\
\mbox{}&-&R_3^{*2} \left(e^{2 h} R_4^{*2} +
e^{2 k_p+2 k_1-2k_x} R_5^{*2} + e^{2 k_p-2 k_1+2 k_x} R_6^{*2} + e^{-2 h}  \right) 
\nonumber  \\
\mbox{}&-&\left. \left(e^{-2 k_p+2 k_1+2 k_x} R_4^{*2} +
e^{-2 h} \left( R_5^{*2} + R_6^{*2} \right)+e^{-2 k_p-2 k_1-2 k_x-4 h}
 \right) \right\}\,, \\
m_B\,=\,\frac{1}{\cal Y} &&\left\{
R_1^{*2} \left(e^{-2 k_p-2 k_1-2 k_x+4 h} R_4^{*2} +
e^{2 h} \left( R_5^{*2} -R_6^{*2} \right)-e^{-2 k_p+2 k_1+2 k_x}
\right) \right.\nonumber \\
\mbox{}&+&R_2^{*2}  \left(e^{2 h} R_4^{*2} +
e^{2 k_p-2 k_1+2k_x} R_5^{*2} - e^{2 k_p+2 k_1-2 k_x} R_6^{*2} -e^{-2 h}  \right)
\nonumber  \\
\mbox{}&+&R_3^{*2} \left(e^{2 h} R_4^{*2} +
e^{2 k_p+2 k_1-2k_x} R_5^{*2} - e^{2 k_p-2 k_1+2 k_x} R_6^{*2} - e^{-2 h}  \right)
\nonumber  \\
\mbox{}&+&\left. \left(e^{-2 k_p+2 k_1+2 k_x} R_4^{*2} +
e^{-2 h} \left( R_5^{*2} - R_6^{*2} \right)-e^{-2 k_p-2 k_1-2 k_x-4 h}
 \right) \right\}\,,\\
m_C\,=\,\frac{1}{\cal Y} &&\left\{
R_1^{*2} \left(e^{-2 k_p-2 k_1-2 k_x+4 h} R_4^{*2} +
e^{2 h} \left( R_5^{*2} +R_6^{*2} \right)+e^{-2 k_p+2 k_1+2 k_x}
\right) \right.\nonumber \\
\mbox{}&-&R_2^{*2}  \left(e^{2 h} R_4^{*2} +
e^{2 k_p-2 k_1+2k_x} R_5^{*2} + e^{2 k_p+2 k_1-2 k_x} R_6^{*2} +e^{-2 h}  \right)
\nonumber  \\
\mbox{}&+&R_3^{*2} \left(e^{2 h} R_4^{*2} +
e^{2 k_p+2 k_1-2k_x} R_5^{*2} + e^{2 k_p-2 k_1+2 k_x} R_6^{*2} + e^{-2 h}  \right)
\nonumber  \\
\mbox{}&-&\left. \left(e^{-2 k_p+2 k_1+2 k_x} R_4^{*2} +
e^{-2 h} \left( R_5^{*2} + R_6^{*2} \right)+e^{-2 k_p-2 k_1-2 k_x-4 h}
 \right) \right\}\,, \\
m_D\,=\,\frac{1}{\cal Y} &&\left\{
R_1^{*2} \left(e^{-2 k_p-2 k_1-2 k_x+4 h} R_4^{*2} +
e^{2 h} \left(-R_5^{*2} +R_6^{*2} \right)-e^{-2 k_p+2 k_1+2 k_x}
\right) \right.\nonumber \\
\mbox{}&+&R_2^{*2}  \left(e^{2 h} R_4^{*2} -
e^{2 k_p-2 k_1+2k_x} R_5^{*2} + e^{2 k_p+2 k_1-2 k_x} R_6^{*2} -e^{-2 h}  \right)
\nonumber  \\
\mbox{}&+&R_3^{*2} \left(e^{2 h} R_4^{*2} -
e^{2 k_p+2 k_1-2k_x} R_5^{*2} + e^{2 k_p-2 k_1+2 k_x} R_6^{*2} - e^{-2 h}  \right)
\nonumber  \\
\mbox{}&+&\left. \left(e^{-2 k_p+2 k_1+2 k_x} R_4^{*2} -
e^{-2 h} \left( R_5^{*2} + R_6^{*2} \right)-e^{-2 k_p-2 k_1-2 k_x-4 h}
 \right) \right\}\,,
\end{eqnarray}
\end{subequations}
\end{widetext}

\noindent where ${\cal Y}$ is the thermodynamic limit of the scaled partition function,

\begin{widetext}
\begin{eqnarray}  \label{espf}
{\cal Y}&=&\lim_{M \rightarrow \infty}\,\frac{{\cal Z}_M}{(Z^a_{--}  Z^b_{--})^2} \,=\,
\nonumber \\
\mbox{}&&
R_1^{*2} \left(e^{-2 k_p-2 k_1-2 k_x+4 h} R_4^{*2} +
e^{2 h} \left( R_5^{*2} +R_6^{*2} \right)+e^{-2 k_p+2 k_1+2 k_x}
\right) \nonumber \\
\mbox{}&+&R_2^{*2}  \left(e^{2 h} R_4^{*2} +
e^{2 k_p-2 k_1+2k_x} R_5^{*2} + e^{2 k_p+2 k_1-2 k_x} R_6^{*2} +e^{-2 h}  \right)
\nonumber  \\
\mbox{}&+&R_3^{*2} \left(e^{2 h} R_4^{*2} +
e^{2 k_p+2 k_1-2k_x} R_5^{*2} + e^{2 k_p-2 k_1+2 k_x} R_6^{*2} + e^{-2 h}  \right)
\nonumber  \\
\mbox{}&+& \left(e^{-2 k_p+2 k_1+2 k_x} R_4^{*2} +
e^{-2 h} \left( R_5^{*2} + R_6^{*2} \right)+e^{-2 k_p-2 k_1-2 k_x-4 h}
 \right)\,.
\end{eqnarray}
\end{widetext}

In order to obtain the Bethe lattice free energy, we follow the Gujrati's argument
[\onlinecite{gujrati95}], as presented by Oliveira {\it et. al}  [\onlinecite{oss09}], obtaining

\begin{widetext}
\begin{equation} \label{efep}
\phi\,=\,\lim_{M \rightarrow \infty}\,-\frac{T}{2} \,\ln{\frac{{\cal Z}_{M+1}}
{{\cal Z}_M^2}}\,=\,-\frac{T}{2} \,\ln{\frac{D_a^2 D_b^2}{{\cal Y}}}\,=\,
-T \left( \ln{D_a} +\ln{D_b}-\frac{1}{2} \ln{{\cal Y}}\right) \,,
\end{equation}
\end{widetext}

\noindent and the first-order transition lines were calculated by equalizing the 
free energies of both phases.



%

\begin{thebibliography}{99}
%
%
\bibitem{Diep}
{\it Frustrated Spin Systems}, 2nd ed., edited by H. T. Diep (World
Scientific, Singapore, 2013)

%
\bibitem{ZitoTrig}
M. V. Gvozdikova, P. E. Melchy and M. E. Zhitomirsky, J. Phys.: Condens. Matter {\bf 23}  164209 (2011)
%
\bibitem{Wannier}
G. H. Wannier, Phys. Rev. {\bf 79} 357 (1950)).
%
\bibitem{OBDkagome} 
J. T. Chalker, P. C. W. Holdsworth  and E. F. Shender, Phys. Rev. Lett. {\bf 68} 855 (1992)
%
\bibitem{GaneshJ1J2}
A. Mulder, R. Ganesh, L. Capriotti, and A. Paramekanti
Phys. Rev. B {\bf 81}, 214419 (2010)
%
\bibitem{RosalesJ1J2}
H. D. Rosales,  D. C. Cabra,  C. A. Lamas, P. Pujol,  and M. E. Zhitomirsky, Phys. Rev. B {\bf 87}, 104402 (2013)
%
\bibitem{Ganesh2011}
R. Ganesh, D. N. Sheng, Y. J. Kim, and A. Paramekanti,
Phys. Rev. B {\bf 83}, 144414 (2011).
%
\bibitem{Kandpal2011}
H. C. Kandpal and J. van den Brink,
Phys. Rev. B {\bf 83}, 140412(R) (2011).
%
\bibitem{Matsuda2010}
M. Matsuda, M. Azuma, M. Tokunaga, Y. Shimakawa, and N. Kumada,
Phys. Rev. Lett. {\bf 105}, 187201 (2010).
%
\bibitem{Lamas2016}
H. Zhang, C. A. Lamas, M. Arlego, and W. Brenig
Phys. Rev. B {\bf 93}, 235150 (2016)
%
\bibitem{Richter2017}
T. Krokhmalskii, V. Baliha, O. Derzhko, J. Schulenburg, and J. Richter
Phys. Rev. B {\bf 95} 094419 (2017)
%
\bibitem{AbInitio2017}
M. Alaei, H.  Mosadeq, I. Abdolhossaini Sarsari and F.  Shahbazi, arXiv:1702.05255
%
\bibitem{Bishop2017}
R. F. Bishop and P. H. Y. Li
Phys. Rev. B {\bf 95}, 134414 (2017), R. F. Bishop and P. H. Y. Li eprint arXiv:1708.06162
%
\bibitem{Smirnova2009}
O. Smirnova {\it et al.},
J. Am. Chem. Soc., {\bf 131}, 8313 (2009);
S. Okubo {\it et al.},
J. Phys.: Conf. Ser. {\bf 200}, 022042 (2010).
%
\bibitem{Albarracin16}
F. A. G\'omez Albarrac\'in and H. D. Rosales
Phys. Rev. B {\bf 93}, 144413 (2016).
%
\bibitem{foot1} The states that are not listed in the table are those obtained from flipping all the spins in the UAF ($4$ states) and U ($1$ state) configurations, 
leading to a negative magnetisation.
%
\bibitem{baxter} 
R.J. Baxter, {\it  Exactly Solved Models in Statistical 
Mechanics}, Academic Press (1982).
%
\bibitem{hio99} 
Chin-Kun Hu, N. Sh. Izmailian, and K. B. Oganesyan, 
Phys. Rev. E {\bf 59}, 6489 (1999).
%
\bibitem{albayrak} O. Canko and E. Albayrak Phys. Rev. E {\bf 75},
011116 (2007); E.  Albayrak and S. Yilmaz, J. Phys. Condensed Matter {\bf 19},
376212 (2007); E. Albayrak, A. Yigit and S. Akkaya, J. Magn. Magn. Mater.
{\bf 310}, 98 (2007);
E. Albayrak, A. Yigit and S. Akkaya, J. Magn. Magn. Mater. {\bf 320}, 2241
(2008); E. Albayrak and S. Akkaya, Physica Scripta {\bf79}, 065005 (2009);
E. Albayrak and  A. Yigit, Acta Phys. Pol. A {\bf 116}, 127 (2009).
%
\bibitem{serra14} P. Serra and J. F. Stilck, J. Stat. Mech. P04002 (2014).
%
\bibitem{gujrati95} P. D. Gujrati, Phys. Rev. Lett. {\bf 74}, 809 (1995).
%
\bibitem{oss09} T. Oliveira, J. F. Stilck and P. Serra, 
Phys. Rev. E {\bf 80}, 041804 (2009). 
%
\bibitem{Metropolis}
N. Metropolis, A. W. Rosenbluth, M. N. Rosenbluth, A. H.
Teller, and E. Teller, J. Chem. Phys. {\bf 21}, 1087 (1953).
%
\bibitem{WangLandau01}
F. Wang and D. P. Landau
Phys. Rev. Lett. {\bf 86}, 2050 (2001)
%
\bibitem{IsingWL2002} 
Y. Okabe, Y. Tomita, C. Yamaguchi, Comput. Phys. Commun. {\bf 146}, 63–68 (2002).
%
\bibitem{Potts2001}
C. Yamaguchi, Y. Okabe, J. Phys. A: Math. Gen. {\bf 34}, 8781 (2001).
%
\bibitem{FM2006}
C. Zhou, T. C. Schulthess, S. Torbrugge and D. P. Landau, Phys. Rev. Lett. {\bf 96}, 120201 (2006).
%
\bibitem{SpinIceFerreyra} 
M. V. Ferreyra , G. Giordano, R. Borzi, J. J. Betouras and S. A. Grigera,  Eur. Phys. J. B {\bf 89} 51 (2016) 
%
\bibitem{Belardinelli07}
R. E. Belardinelli, and V. D. Pereyra
Phys. Rev E {\bf 75}, 046701 (2007)
%
\bibitem{LandauFreeE} 
J. C. Tol\'edano and  P. Tol\'edano: The Landau Theory of Phase Transitions. World Scientific publishing Co. Pte. Ltd., Singapore (1987);
K. Watanabe and S. Munetaka,  J. Phys. Soc. Jpn. {\bf 80}, 093001 (2011).
%
\bibitem{foot2} We here note that although in theory this may be easily extended to include more that one $\text{OP}$, in practice to ensure the convergence of the 
algorithm the density of states will be calculated as a function of up to two variables.
%
\bibitem{foot3} Due the symmetry $J_1\leftrightarrow J_x$ in the Hamiltonian 
(Eq. (\ref{eq:Hamiltonian})), a similar phase diagram can be obtained  for the case $J_1/J_p$ vs $T/J_p$, with $J_x=J_p$ changing AF$_2$ $\to$ U$_2$
%
\bibitem{Pathria}
R. K. Pathria and P. D. Beale. {\it Statistical Mechanics} Elsevier Science, Amsterdam, 1996.
%
\bibitem{Bramwell01}
S. T. Bramwell and M. J. P. Gingras
Science {\bf 294}, 1495 (2001).
%
\bibitem{Harris97}
M. J. Harris, S. T. Bramwell, D. F. McMorrow, T. Zeiske, and K. W. Godfrey
Phys. Rev. Lett. {\bf 79}, 2554 (1997).
%
\bibitem{Bramwell01_2}
S. T. Bramwell, M. J. Harris, B. C. den Hertog, M. J. P. Gingras,
J. S. Gardner, D. F. McMorrow, A. R. Wildes, A. L. Cornelius,
J. D. M. Champion, R. G. Melko, and T. Fennell
Phys. Rev. Lett. {\bf 87}, 047205 (2001).
%
\bibitem{Ramirez99}
A. P. Ramirez, A. Hayashi, R. J. Cava, R. Siddharthan, and B. S. Shastry
Nature (London) {\bf 399}, 333 (1999).
%
\bibitem{Fennell02}
T. Fennell, O. A. Petrenko, G. Balakrishnan, S. T. Bramwell, J. D.
M. Champion, B. F{\aa}k, M. J. Harris, and D. M. Paul
Appl. Phys. A {\bf 74}, 889 (2002).
%
\bibitem{Zvyagin}
A. A. Zvyagin, Low Temperature Physics  {\bf 39}, 1159 (2013)
%
\bibitem{LFE_comment} 
K. Binder, Philosophical Magazine Letters Vol. {\bf 87} , Iss. 11, (2007). 
%
\bibitem{Villain89}
J. Villain, R. Bidaux, J. P. Carton, and R. Conte,
J. Phys. (Paris) {\bf 41}, 1263 (1980).
%
\bibitem{Shender82}
E. F. Shender, Sov. Phys. JETP {\bf 56}, 178 (1982).
%
\end{thebibliography}
\end{document}